\documentclass[acmlarge]{acmart}

\AtBeginDocument{%
  }

\setcopyright{cc}
\setcctype{by}
\acmJournal{IMWUT}
\acmYear{2025} \acmVolume{9} \acmNumber{3} \acmArticle{99} \acmMonth{9} \acmPrice{}\acmDOI{10.1145/3749469}

\usepackage{xcolor}
\usepackage{tablefootnote}

\newcommand{\name}{Grab-n-Go}

\usepackage{color,soul}
\usepackage{booktabs} 
\usepackage{subfig} 
\usepackage{graphicx}

\usepackage[utf8]{inputenc}
\usepackage[T1]{fontenc}

\usepackage{siunitx}
\newcommand{\hide}[1]{}
\definecolor{CAT-comment}{rgb}{0.95, 0.2, 0.8}

\definecolor{CJ-color}{RGB}{23, 225, 225}

\sethlcolor{yellow}
\ifdefined\scifihidecomments
    \newcommand{\cheng}[1] {}
    \newcommand{\hyunc}[1] {} 
    \newcommand{\yax}[1] {} 
    \newcommand{\tuoc}[1] {} 
    \newcommand{\sony}[1] {} 
    \newcommand{\ReviewerFeedback}[1] {}  
    \newcommand{\fran}[1] {}
    \newcommand{\rz}[1]{}
    \newcommand{\lw}[1] {}
    \newcommand{\mose}[1] {} 
    \newcommand{\ke}[1] {} 

\else
    \definecolor{burntorange}{rgb}{0.8, 0.33, 0.0}
    \definecolor{cadmiumgreen}{rgb}{0.0, 0.42, 0.24}
    \definecolor{cobalt}{rgb}{0.0, 0.28, 0.67}
    \definecolor{amber}{rgb}{1.0, 0.75, 0.0}
    \definecolor{fashionfuchsia}{rgb}{0.96, 0.0, 0.63}
    \definecolor{brightcerulean}{rgb}{0.11, 0.67, 0.84}
    \definecolor{frenchblue}{rgb}{0.0, 0.45, 0.73}
    \definecolor{darkslateblue}{rgb}{0.28, 0.24, 0.55}
    \definecolor{cerulean}{rgb}{0.0, 0.48, 0.65}
    \definecolor{darkpastelgreen}{rgb}{0.01, 0.75, 0.24}

    \newcommand{\hyunc}[1] { \textcolor{burntorange}{[{\hl{hyunc:}} {#1}}]}
    \newcommand{\yax}[1] { \textcolor{magenta}{[{\hl{yaxuan:}} {#1}]}}
    \newcommand{\tuoc}[1] { \textcolor{darkpastelgreen}{[{\hl{tuochao:}} {#1}]}}
    \newcommand{\sony}[1] { \textcolor{blue}{[{\hl{songyun:}} {#1}]}}
    \newcommand{\ReviewerFeedback}[1] { \textcolor{brightcerulean}{[{Reviewer Feedback:} {#1}}]}
    \newcommand{\fran}[1]{\textcolor{burntorange}{[{francois:}{#1}}]}
    \newcommand{\rz}[1]{\textcolor{teal}{[{Ruidong: }{#1}]}}
    \newcommand{\lw}[1]{\textcolor{fashionfuchsia}{[{liuwei:}{#1}}]}
    \newcommand{\mose}[1]{\textcolor{burntorange}{[{mose:}{#1}}]}
    \newcommand{\ke}[1] { \textcolor{red!55!yellow}{[{Ke:} {#1}}]}

\fi

\ifdefined\scifiblind
    \newcommand{\blind}[1]{[omitted for blind review]}
\else
    \newcommand{\blind}[1]{#1} 
\fi

\newcommand{\etal}{et al.~}

\begin{document}

\title{\name{}: On-the-Go Microgesture Recognition with Objects in Hand}

\author{Chi-Jung Lee}
\email{cl2358@cornell.edu}
\orcid{0000-0002-1887-4000}
\affiliation{
  \institution{Cornell University}
  \city{Ithaca}
  \state{New York}
  \country{USA}
}

\author{Jiaxin Li}
\email{jl2726@cornell.edu}
\orcid{0009-0000-7635-6749}
\affiliation{
  \institution{Cornell University}
  \city{Ithaca}
  \state{New York}
  \country{USA}
}

\author{Tianhong Catherine Yu}
\email{ty274@cornell.edu}
\orcid{0000-0002-3742-0178}
\affiliation{
  \institution{Cornell University}
  \city{Ithaca}
  \state{New York}
  \country{USA}
}

\author{Ruidong Zhang}
\email{rz379@cornell.edu}
\orcid{0000-0001-8329-0522}
\affiliation{
  \institution{Cornell University}
  \city{Ithaca}
  \state{New York}
  \country{USA}
}

\author{Vipin Gunda}
\email{vg245@cornell.edu}
\orcid{0009-0000-5500-2183}
\affiliation{
  \institution{Cornell University}
  \city{Ithaca}
  \state{New York}
  \country{USA}
}

\author{François Guimbretière}
\email{fvg3@cornell.edu}
\orcid{0000-0002-5510-6799}
\affiliation{%
  \institution{Cornell University}
  \city{Ithaca}
  \state{New York}
  \country{USA}
}

\author{Cheng Zhang}
\email{chengzhang@cornell.edu}
\orcid{0000-0002-5079-5927}
\affiliation{%
  \institution{Cornell University}
  \city{Ithaca}
  \state{New York}
  \country{USA}
}

\renewcommand{\shortauthors}{Lee et al.}

\begin{abstract}
As computing devices become increasingly integrated into daily life, there is a growing need for intuitive, always-available interaction methods --- even when users’ hands are occupied. In this paper, we introduce \name{}, the first wearable device that leverages active acoustic sensing to recognize subtle hand microgestures while holding various objects. Unlike prior systems that focus solely on free-hand gestures or basic hand-object activity recognition, \name{} simultaneously captures information about hand microgestures, grasping poses, and object geometries using a single wristband, enabling the recognition of fine-grained hand movements occurring within activities involving occupied hands. A deep learning framework processes these complex signals to identify 30 distinct microgestures, with 6 microgestures for each of the 5 grasping poses. In a user study with 10 participants and 25 everyday objects, \name{} achieved an average recognition accuracy of 92.0\%. A follow-up study further validated \name{}'s robustness against 10 more challenging, deformable objects. These results underscore the potential of \name{} to provide seamless, unobtrusive interactions without requiring modifications to existing objects. The complete dataset, comprising data from 18 participants performing 30 microgestures with 35 distinct objects, is publicly available at \href{https://github.com/cjlisalee/Grab-n-Go_Data}{https://github.com/cjlisalee/Grab-n-Go\_Data} with the DOI: \href{https://doi.org/10.7298/7kbd-vv75}{https://doi.org/10.7298/7kbd-vv75}.

\end{abstract}

\ccsdesc[500]{Human-centered computing~Interaction devices}

\keywords{Wearable, Acoustic Sensing, Smartwatch, Gesture}

\begin{teaserfigure}
  \includegraphics[width=1\textwidth]{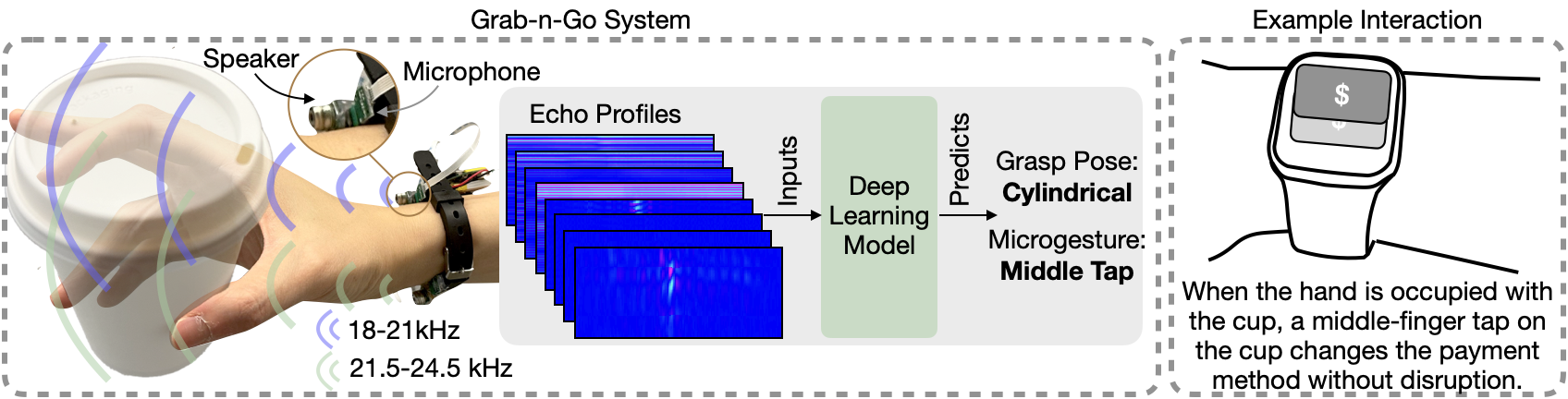}
  \caption{We propose \name{}, a wristband that can recognize 30 microgestures across 35 various everyday objects. With two speaker-microphone pairs on each side of the wrist emitting and receiving the acoustic waves ranging 18-21 kHz and 21.5-24.5 kHz, respectively, echo profiles can be created to infer the microgestures using a customized deep learning model. \name{} enables gestural control when the hands are occupied.}
  \Description{
  This figure shows how the Grab-n-Go System works in a real-life setting, using sound waves to detect hand movements and gestures while holding objects, like a cup. The figure is divided into two parts: the Grab-n-Go System and the Example Interaction.
  On the left side: A hand is shown holding a cup while wearing the Grab-n-Go system on the wrist. The system has two main parts: a speaker and a microphone attached to the wristband. The speaker sends out sound waves in two frequency ranges: 18-21 kHz (shown as purple lines) and 21.5-24.5 kHz (shown as green lines). These sound waves bounce off the hand and the cup, and the microphone captures the echoes (sound reflections). These echoes are shown as stacked blue-colored images in the middle of the left section, called echo profiles. These profiles are fed into a deep learning model. The model looks at the patterns in the echoes to predict the hand’s grasp type (in this case, “Cylindrical”) and the gesture being made (in this case, “Middle Tap”).
  On the right side: The Example Interaction shows how this system can be used. In this example, when the user taps the cup with their middle finger, the system triggers a change in the payment method on a smartwatch without the user needing to let go of the cup or use their other hand. This shows how the system allows the user to interact with technology while still holding an object, making everyday tasks easier.
  }
  \label{fig:teaser}
\end{teaserfigure}

\maketitle
\section{Introduction}
As computing devices become increasingly integrated into daily life, the need for always-available, unobtrusive interaction methods continues to grow. While hands are the primary means of interaction with these devices --- through actions like typing, swiping, and gesturing --- in everyday scenarios, hands are frequently occupied with holding, carrying, or using objects, making traditional input methods impractical. This presents a challenge, as the need to interact with computing devices persists, particularly for quick actions like answering phone calls or controlling smart home devices. 

Subtle hand gestures, or microgestures, performed while holding objects offer a promising solution for always-available input \cite{sharma2019grasping, sharma2021solofinger}. However, recognizing these microgestures in the presence of objects introduces significant challenges. One approach involves instrumenting objects with sensors \cite{schmitz2015capricate, zhang2017electrick}, but this can be costly and not feasible for everyday objects, especially consumables. Wearable-based methods can potentially overcome this limitation by eliminating the need for object modifications \cite{saponas2009enabling, rudolph2022sensing, kim2023vibaware, sharma2023sparseimu}. Yet, existing wearable solutions typically focus solely on tracking hand movements, ignoring the objects themselves. This limitation reduces their generalizability across various objects. As a result, most prior work has been restricted to recognizing microgestures across a relatively small set of objects, typically around 10 or fewer.


Recent advancements in wearable computing have demonstrated the potential of active acoustic sensing for tracking fine-grained movements of different body parts \cite{li2022eario, mahmud2023posesonic, zhang2023hpspeech, sun2023echonose, zhang2023echospeech, li2024eyeecho, mahmud2024munchsonic, parikh2024echoguide, li2024sonicid, mahmud2024actsonic}. More specifically, while previous work has successfully captured fine-grained hand movements and some hand-object interactions \cite{lee2024echowrist, yu2024ring, lim2025spellring}, these systems primarily focus on free-hand gestures or the identification of broader object-related activities. Notably, the recognition of hand microgestures performed while holding objects remains an underexplored area within this domain.

The presence of objects in hand introduces unique challenges for active acoustic sensing:

\begin{enumerate}

    \item \textbf{Diverse Object Geometries}: Varying object shapes lead to different grasping poses and signal variations. The challenge is to build a system that can generalize microgesture recognition across a wide range of object geometries.
    
    \item \textbf{Occlusion}: Objects in hand can block sensor views, interfering with signal capture --- especially when detecting subtle hand and finger movements. The challenge is to reliably capture and distinguish these subtle finger motions despite significant acoustic signal reflections caused by various object shapes and grasping poses. 

\end{enumerate}

Despite these challenges, we hypothesize that air-borne active acoustic sensing can simultaneously capture information about hand-grasping poses, object geometries, and microgestures using a single wristband. This approach offers a promising solution for robust microgesture recognition across a wide range of objects without requiring modifications to them.

This paper explores the following research question: 

\begin{itemize}
    \item How can we develop a wristband with active acoustic sensing that reliably recognizes a rich set of microgestures while a user holds various objects in different grasping poses?
\end{itemize}

To answer this question, we present \name{}, a wristband-based system designed to recognize 30 hand microgestures while holding various objects. \name{} employs active acoustic sensing by using two embedded speakers to emit inaudible acoustic waves toward the hand and the object. Two microphones on the wristband capture the reflected waves, which form unique patterns corresponding to each microgesture.

We evaluated \name{} in a user study with 10 participants. The study examined 5 distinct grasping poses, defined by Schlesinger's grasp taxonomy~\cite{sharma2019grasping}. Each of these grasping poses involved 6 microgestures, resulting in a total of 30 microgestures (Fig. \ref{fig:objects-gestures}). To ensure generalizability, we included 5 diverse objects per grasping pose, totaling 25 objects. Each participant performed 6 microgestures 24 times for 2 randomly assigned objects within each grasping pose, resulting in 1,440 microgesture samples per participant. Our model achieved an average accuracy of 92.0\% across all objects, demonstrating the effectiveness of \name{} in recognizing hand microgestures while hands are occupied. 

To further explore system boundaries, we conducted a follow-up study with 8 additional participants specifically targeting 10 more challenging, deformable objects. These objects introduce greater variability across testing sessions and continuous shape transformations during microgesture execution. When evaluating on a combined dataset from both studies, and assessing performance on one object per grasping pose (compared to two in the initial study), we observed recognition accuracies of 95.0\% for non-deformable objects and 92.9\% for deformable objects. This suggests that \name{} benefits from a larger and more diverse training dataset and exhibits promising capability in handling the increased complexity introduced by deformable objects. The complete dataset with 18 participants and 35 objects is publicly available at \href{https://github.com/cjlisalee/Grab-n-Go_Data}{https://github.com/cjlisalee/Grab-n-Go\_Data} with the DOI: \href{https://doi.org/10.7298/7kbd-vv75}{https://doi.org/10.7298/7kbd-vv75}.

In summary, \name{} is the first wearable sensing device specifically designed to recognize a rich set of microgestures while holding various objects, paving the way for more natural and always-available hand interactions. We conclude with a discussion of the challenges and design implications for future wearable systems.

The contributions of this paper are: 

\begin{itemize}
  \item We show that active acoustic sensing on a wristband can effectively recognize hand microgestures even when users are holding various objects.

  \item We conducted a user study with 10 participants performing 30 microgestures across 5 distinct grasping poses and 25 different objects. This was followed by another study with 8 participants using 10 deformable objects. The two studies validated the system's robustness and generalizability.

  \item We released the dataset, comprising a total of 20,160 microgesture instances performed by 18 participants across 35 distinct objects, to facilitate further research in this domain.

  \item We provide practical design implications for future wearable devices, outlining the challenges and opportunities for enabling seamless, always-available hand interactions even when hands are occupied.
\end{itemize}

\section{Related Work}

\begin{table*}[ht]
    \centering
    \caption{Comparison with Prior Work}
    \resizebox{\textwidth}{!}{
        \begin{tabular}{|c|c|c|c|c|c|c|c|c|}
            \hline
             & \textbf{Technique} & \textbf{Algorithm} & \textbf{Form Factor} & \textbf{Gestures} & \textbf{Objects} & \textbf{Performance} & \textbf{Remounting} \\
            \hline
            Saponas et al. \cite{saponas2009enabling} & EMG & SVM & Armband & 4 & 2 & 85\% (tumbler) & $\times$ \\
             &  &  &  & (Finger Press) &  & 88\% (bag) &  \\
            \hline
            Rudolph et al. \cite{rudolph2022sensing} & Capacitive Sensing & LDA & Wristband & 6 & 6 & 99\% & $\times$ \\
             &  &  &  & (Object Interaction) &  &  \\
            \hline
            VibAware \cite{kim2023vibaware} & Active \& Passive & SVM & Wristband & 12 & 4 & 85.7\% & $\times$\\
             & Acoustic Sensing &  & + Ring &  & (3D-Printed Prop) &  &  \\
            \hline
            SparseIMU \cite{sharma2023sparseimu} & Finger Joint IMUs & RF & On-Skin & 19 & 12 & 0.93 F1 Score & \checkmark \\
            \hline
            \textbf{Grab-n-Go} & \textbf{Active Acoustic Sensing} & \textbf{ResNet} & \textbf{Wristband} & \textbf{30} & 25 & 92\% & \checkmark \\
            \hline
        \end{tabular}
        }
    \label{tab:comparison_table}
\end{table*}


Sensing hand microgestures while holding objects is challenging due to the subtle nature of these gestures and the wide variety of objects the hand may engage with. Prior research mainly focuses on two approaches, depending on sensor placement: instrumenting the objects or instrumenting the hands. Approaches that instrument the objects integrate sensors directly into the objects themselves, enabling them to detect microgestures, while approaches that instrument the hands rely on wearable sensors placed on the hand to capture microgestures. In this section, we explore each of these approaches in detail and discuss their relationship to our proposed \name{} system.

\subsection{Gesture Sensing Instrumenting the Objects}
To enable microgesture sensing on objects, researchers have explored embedding sensors directly into various objects. One approach leverages the inherent conductivity of objects for touch sensing by connecting them to sensor boards \cite{sato2012touche, zhang2017electrick}. However, since many everyday objects are not naturally conductive, alternative methods have been developed, such as applying touch sensors to the surface \cite{savage2012midas, pourjafarian2022print, kawahara2013instant, zhang2017electrick, wessely2020sprayable} or embedding them inside the objects themselves \cite{palma2024capacitive, tejada2020airtouch}. To automate the fabrication process, some researchers have focused on integrating touch sensors during manufacturing. Capricate \cite{schmitz2015capricate} and MetaSense \cite{gong2021metasense} enable touch sensing on 3D-printed objects by incorporating capacitive touch sensors during the 3D-printing process. While these approaches enable gestural user interfaces on various objects and support object-oriented interactions, they face a significant limitation: the need to instrument every object. This requirement presents a significant challenge for widespread adoption, as it is impractical to instrument all the objects people encounter in daily life.

To tackle the challenge, researchers have explored instrumenting more ubiquitous objects or materials. One example is cords, which are commonly found in everyday life --- such as jacket drawstrings, charging cables, and bracelets. By integrating sensors into cords, these everyday objects can be enhanced with sensing capabilities \cite{shahmiri2019serpentine, chen2022imperceptible, olwal2018braid, olwal2020textile, ku2020threadsense}. In addition, textiles, which are prevalent in various aspects of daily life, from furniture to clothing, have been proposed as a key medium for ubiquitous computing. E-textiles, in particular, have shown significant potential for gesture recognition by utilizing fabric-based sensors to detect hand movements \cite{yu2023uknit, wu2020fabriccio, wu2021project}. However, despite the potential of these innovations, the number of everyday objects people frequently use still far exceeds those that could be instrumented.

Instrumenting every object for microgesture sensing presents challenges. It is impractical for consumables and widely used everyday objects, as it would require constant modification and maintenance. Furthermore, augmenting objects with embedded sensors often needs significant customization, which can be complex, costly, and may not scale effectively across diverse objects. In contrast, instrumenting the hand with a wearable device eliminates the need to alter objects, offering a more flexible and ubiquitous solution for recognizing hand gestures while holding various objects. By focusing on the user rather than the object, wearable-based approaches enable seamless and consistent microgesture recognition across different scenarios. In the next subsection, we explore gesture-sensing approaches that instrument the hands.


\subsection{Gesture Sensing Instrumenting the Hands}
Given watches' long-standing social acceptance and minimal disruption to daily routines, wrist-worn devices, which can potentially be integrated into watches, or smartwatches, have emerged as a prime focus. In this subsection, we focus on wrist-worn devices for hand gesture recognition.


Recognizing hand gestures \cite{de2020real, du2017semi, kerber2017user, raurale2018emg, kim2022ether, zhang2015tomo, zhang2016advancing, mcintosh2017echoflex, zhang2018fingerping, truong2018capband, hsiao2023demo, park2011gesture, zhang2016watchout,
zhang2016tapskin, li2018wristwash} or even continuously tracking the hand poses \cite{liu2021wr, liu2021neuropose, kyu2024eitpose, wu2020back, kim2012digits, yeo2019opisthenar, lee2024echowrist, hu2020fingertrak, zhang2017soundtrak} has been extensively studied. However, most of these works are confined to free-hand gestures and do not investigate how the system performs when hands are occupied by objects.
The presence of objects in the hand can introduce occlusion and noise, hindering gesture recognition. For instance, FingerTrak \cite{hu2020fingertrak} experienced significant performance degradation in hand tracking when the hand is holding small objects, underlining the challenges posed by in-hand objects.

Some other methods have been proposed to recognize hand gestures with objects in hand. Leveraging forearm electromyography (EMG), Saponas et al. \cite{saponas2009enabling} developed an armband capable of distinguishing pinching movements 
when holding a tumbler or a handbag. Their approach utilized features extracted from filtered 1D EMG signals to train a support vector machine (SVM) classifier. Rudolph et al. \cite{rudolph2022sensing} presented a wristband that can recognize force, grasp, and object manipulation based on capacitive sensing. This system extracted statistical features from 2D heatmaps representing the spatial distribution of capacitive magnitude signals and trained linear discriminant analysis (LDA) models. VibAware \cite{kim2023vibaware} supports tap and swipe when grasping different shapes of objects using bio-acoustic sensing. Their method extracted features from 1D filtered bio-acoustic signals to train an SVM classifier. SparseIMU \cite{sharma2023sparseimu} is a platform that supports the required IMUs on hand for sensing different gesture sets, employing features extracted from 1D filtered IMU signals to train a random forest (RF) classifier.

However, these sensing approaches focus mostly on the internal status of the hands yet lack information on the objects as well as the interaction between hands and objects, i.e., grasping poses. As a result, these works mostly investigate a small number of objects that have similar shapes or sizes. The ability of these systems to function across various objects has not been extensively investigated. In contrast, \name{} leverages active acoustic sensing, which captures information about not only the hands but also the objects around them. Therefore, \name{} can consistently work across various everyday objects. Notably, the rich spatiotemporal information contained in our 2D feature maps generated by our approach, the echo profiles, presents greater complexity than the signals processed in prior work, necessitating more sophisticated approaches beyond the generic machine learning methods employed in previous work. Our analysis demonstrates that advanced deep learning architectures more effectively extract and leverage the complex patterns inherent in these acoustic feature maps and generate better results.

The closest work to ours is EchoWrist \cite{lee2024echowrist}, which employs active acoustic sensing on a wristband for tracking free-hand poses and recognizing basic hand-object interactions. However, as discussed previously, recognizing subtle hand gestures while holding objects presents unique challenges, including grasping variations and signal occlusions, which have not been addressed in prior work. While EchoWrist can identify broader hand-object activities. e.g., holding chopsticks or stirring with chopsticks, it does not capture the fine-grained hand movements occurring within these activities. To the best of our knowledge, \name{} is the first wearable device to demonstrate the use of active acoustic sensing for recognizing a rich set of hand microgestures while holding various objects, showing clear improvements over existing microgesture recognition approaches for occupied hands.

\section{Design Considerations}

To address the research question: \textbf{"How can we develop a wristband with active acoustic sensing that reliably recognizes a rich set of microgestures while a user holds various objects in different grasping poses?"}, and ultimately enable always-available gestural input in daily life while minimizing disruption to ongoing activities, we propose the following design considerations:

\subsection{Microgesture Design and Generalization across Various Objects}\label{sec:object-gesture-set}

\begin{figure*}[ht]
  \centering
  \includegraphics[width=1\textwidth]{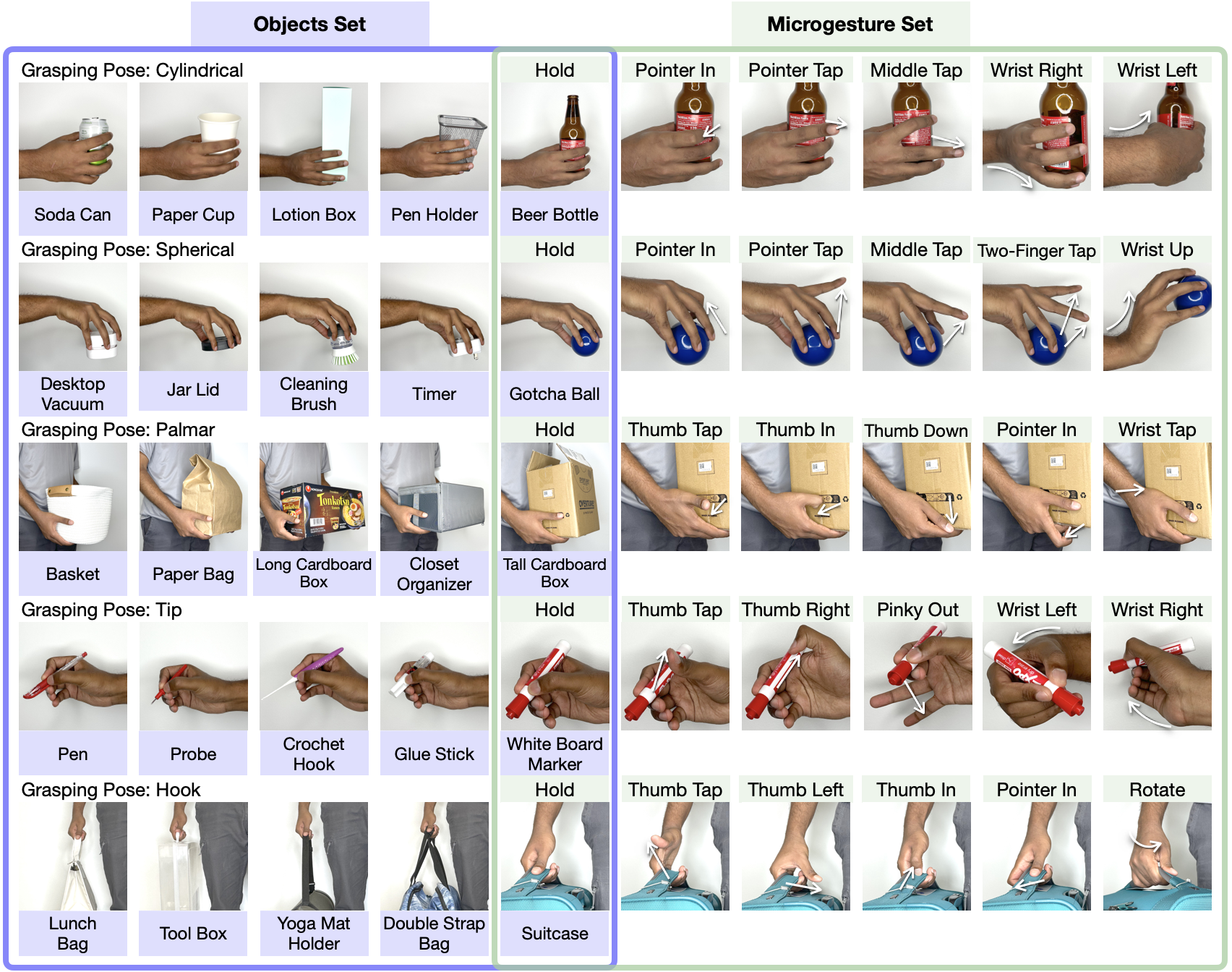}
  \caption{The object and microgesture sets used for evaluating the capability of \name{}.}
  \Description{
  This figure shows different types of hand grasps and gestures used to interact with objects. The figure is divided into five rows, and each row represents a different grasp type. On the left side of each row (boxed in purple), there are objects, and on the right side (boxed in green), there are hand gestures.
  Top Row (Cylindrical Grasp): This row shows five cylindrical objects—Soda Can, Paper Cup, Lotion Box, Pen Holder, and Beer Bottle—positioned from left to right. Next to each object are six hand gestures: “Hold,” “Pointer In,” “Pointer Tap,” “Middle Tap,” “Wrist Right,” and “Wrist Left.” These gestures focus on grasping round objects using a full-hand grip.
  Second Row (Spherical Grasp): The second row features round-shaped items—Desktop Vacuum, Jar Lid, Cleaning Brush, Timer, and Gotcha Ball. The six hand gestures associated with these objects are “Hold,” “Pointer In,” “Pointer Tap,” “Middle Tap,” “Two-Finger Tap,” and “Wrist Up.” These gestures emphasize finger positioning and wrist movements for spherical items.
  Middle Row (Palmar Grasp): In the middle row, larger objects such as a Basket, Paper Bag, Long Cardboard Box, Closet Organizer, and Tall Cardboard Box are depicted. The gestures for these items include “Hold,” “Thumb Tap,” “Thumb In,” “Thumb Down,” “Pointer In,” and “Wrist Tap.” This set focuses on gripping objects with the palm, with movements mainly centered around the thumb and wrist.
  Fourth Row (Tip Grasp): This row presents smaller or thinner objects like a Pen, Probe, Crochet Hook, Glue Stick, and Whiteboard Marker. The corresponding gestures are “Hold,” “Thumb Tap,” “Thumb Right,” “Pinky Out,” “Wrist Left,” and “Wrist Right,” highlighting precision grips using the fingertips.
  Bottom Row (Hook Grasp): The final row displays objects that are carried by hooking the hand --- Lunch Bag, Tool Box, Yoga Mat Holder, Double Strap Bag, and Suitcase. The gestures used for these objects are “Hold,” “Thumb Tap,” “Thumb Left,” “Thumb In,” “Pointer In,” and “Rotate.” These gestures involve strong grips where the hand hooks around the object.
  }
  \label{fig:objects-gestures}
\end{figure*}

In everyday lives, people's hands are frequently occupied by a wide range of objects, which vary significantly in shape, material, size, and weight \cite{sharma2024graspui, zheng2011investigation}. This inherent diversity poses a significant challenge for consistently recognizing the same microgesture across different objects. Generalizing microgesture recognition to work effectively across such a wide variety of objects remains a key research problem.

Prior research \cite{sharma2019grasping, Schlesinger1919, mackenzie1994grasping} suggests that microgestures are largely determined by the grasping poses and the geometry of the objects being held. This indicates that even when objects differ in shape, material, size, or weight, they may still allow for the same set of microgestures if they are held using similar grasping poses. In other words, the way human hands hold an object plays a more critical role in determining microgestural possibilities than the object's specific characteristics. By prioritizing grasping poses rather than individual object properties, the system can generalize microgesture recognition across a wide range of objects. This approach significantly reduces the need for extensive object-specific training data, enhancing scalability and adaptability. Instead of training the system on a vast number of objects, it can focus on a finite set of grasping poses that naturally emerge across different interactions, making microgesture recognition more efficient and robust in diverse real-world scenarios.

To ensure generalizability across various objects, \name{} aims to classify the same set of microgestures when performed with the same grasping pose regardless of the specific object being held. This approach assumes that for all objects typically grasped in a given pose, the available microgestures remain consistent. By focusing on grasping poses rather than object-specific characteristics, \name{} can reduce the need for extensive object-specific training data and enhance adaptability across a diverse range of everyday objects. Following prior research \cite{kim2023vibaware, sharma2023sparseimu, saponas2009enabling, fan2018your}, we adopt the grasping poses classification system defined by Schlesinger \cite{Schlesinger1919} as the basis for grouping microgestures. However, we exclude the “lateral” grasp, as Sharma~\etal~\cite{sharma2019grasping} point out that this grasping pose constrains the most dexterous fingers --- the thumb and pointer finger --- making it challenging to perform microgestures with the remaining fingers. By focusing on grasping poses that allow for finer finger movements, we optimize microgesture sets while maintaining broad applicability across different everyday scenarios.

Our microgesture set design (Fig. \ref{fig:objects-gestures}) is informed by prior research on microgestures performed with occupied hands \cite{kim2023vibaware, sharma2019grasping, rudolph2022sensing}. We define a set of 5 dynamic microgestures for each grasping pose. These microgestures are carefully selected based on their feasibility, distinctiveness, and potential for seamless execution while holding various objects. In addition to the dynamic microgestures, we incorporate a static holding state, in which the user simply holds an object without other movements. This state serves as a neutral baseline, preventing unintended activations and enabling practical, real-world applications where microgesture input should only be recognized when explicitly performed. We call it the Hold microgesture in the rest of the paper. In total, our microgesture set consists of 30 unique microgestures, covering a diverse range of grasping poses and objects. The 5 grasping poses and their 6 corresponding microgestures are as follows:


\begin{itemize}
    \item Cylindrical: 
    \begin{itemize}
        \item Grasping Pose: An open fist grip for cylindrical objects (\textit{e.g.}, paper cups) where the thumb is positioned on one side of the object and the other four fingers on the opposite side. 
        \item Constraints: The thumb and at least two fingers are required to maintain the grip, with the pointer and middle fingers having greater dexterity. Therefore, the proposed microgestures involve movements of the pointer finger, middle finger, and wrist.
        \item Proposed Microgestures: Hold, Pointer In, Pointer Tap, Middle Tap, Wrist Right, and Wrist Left.
    \end{itemize}
    \item Spherical: 
    \begin{itemize}
        \item Grasping Pose: An open fist grip for spherical objects (\textit{e.g.}, tennis balls), where all the fingers are evenly distributed around the objects. 
        \item Constraints: The thumb and two additional fingers are essential for maintaining the hold. To avoid altering the grasping position while performing microgestures, movements primarily involve the pointer finger, middle finger, and wrist. 
        \item  Proposed Microgestures: Hold, Pointer In, Pointer Tap, Middle Tap, Two-Finger Tap, and Wrist Up.
    \end{itemize}
    \item Palmar: 
    \begin{itemize}
        \item Grasping Pose: A posture for holding flat, thick objects (\textit{e.g.}, moving boxes), where the thumb stabilizes the object from the side while the other four fingers support it from underneath. 
        \item Constraints: To avoid dropping the object and consider finger dexterity, we propose microgestures using the thumb, pointer finger, and wrist.
        \item  Proposed Microgestures: Hold, Thumb Tap, Thumb In, Thumb Down, Pointer In, and Wrist Tap.
    \end{itemize}
    \item Tip: 
    \begin{itemize}
        \item Grasping Pose: A grip for sharp and small objects (\textit{e.g.}, pens), requiring at least two fingers for secure handling. 
        \item Constraints: The ring and pinky fingers have the most freedom of movement, though the ring finger is difficult to move independently. As a result, the proposed microgestures focus on the thumb, pinky finger, and wrist. 
        \item Proposed Microgestures: Hold, Thumb Tap, Thumb Right, Pinky Out, Wrist Left, and Wrist Right.
    \end{itemize}
    \item Hook:
    \begin{itemize}
        \item  Grasping Pose: A posture for carrying heavy objects with handles (\textit{e.g.}, suitcase handles) where all fingers except the thumb are used to hook and secure the object.  
        \item Constraints: The grip provides the most flexibility for thumb movements. 
        \item Proposed Microgestures: Hold, Thumb Tap, Thumb Left, Thumb In, Pointer In, and Rotate.
    \end{itemize}
\end{itemize}

To create a seamless and intuitive user experience, we designed all microgestures to start from the natural holding position and return to this initial state upon completion. For example, if the user is holding a tumbler and wants to perform the Pointer In microgesture, the user simply slides their pointer finger inward and then returns it to its original position. There is no need to first reposition the finger before executing the gesture, reducing cognitive effort and making interactions more fluid and natural.

\subsection{The Choice of Sensing Technique and Form Factor}    \label{sec:choice-of-technique}
Hands are frequently occupied with objects during daily activities, presenting a significant challenge for wearable sensing systems. The presence of objects can obstruct sensors, leading to signal occlusion or undesired interference. To address this challenge, a sensing technique is required that not only mitigates the effects of occlusion but also leverages the unique signal characteristics introduced by object interference to enhance recognition. Ideally, the sensing technique should capture comprehensive information about grasping poses, hand microgestures, and object geometry simultaneously. 

In addition, for microgesture recognition to be practical in everyday settings, factors such as cost, power efficiency, and user comfort must be carefully considered. An ideal sensing solution should adopt a widely accepted and minimally obtrusive form factor while integrating affordable, readily available sensors with low power consumption. By balancing accuracy, practicality, and usability, the system can support seamless and unobtrusive interaction in real-world scenarios. 

Considering these factors, we propose \name{}, a wristband embedded with active acoustic sensing. Wristbands, which can be seamlessly integrated into smartwatches, have long been one of the most popular and widely accepted wearable form factors, offering a comfortable and unobtrusive user experience. More importantly, active acoustic sensing employs compact, low-cost sensors and has demonstrated promising results in estimating hand poses when hands are empty \cite{lee2024echowrist, yu2024ring}. With airborne acoustic signals, both the hand and the in-hand objects reflect the emitted waves, encoding information about their geometry. When a user performs microgestures, the resulting signal changes form distinct patterns that are independent of the specific object being held. This enables \name{} to capture a rich set of information, including object shape, grasping poses, and microgestures, all based on a single sensing modality. To the best of our knowledge, no existing system has explored this approach or achieved such comprehensive sensing capabilities. By jointly learning grasping poses, object properties, and microgestures, \name{} has the potential to recognize a wide range of hand microgestures across various objects --- using only a single wristband.

\section{The Design and Implementation of \name{}}
To recognize the microgestures while hands are occupied with various objects, we designed a compact wristband powered by active acoustic sensing and customized machine-learning inference pipelines. In this section, we provide a detailed overview of \name{}'s hardware design, sensing principle, and machine learning pipeline.

\begin{figure*}[t]
  \centering
\includegraphics[width=1\textwidth]{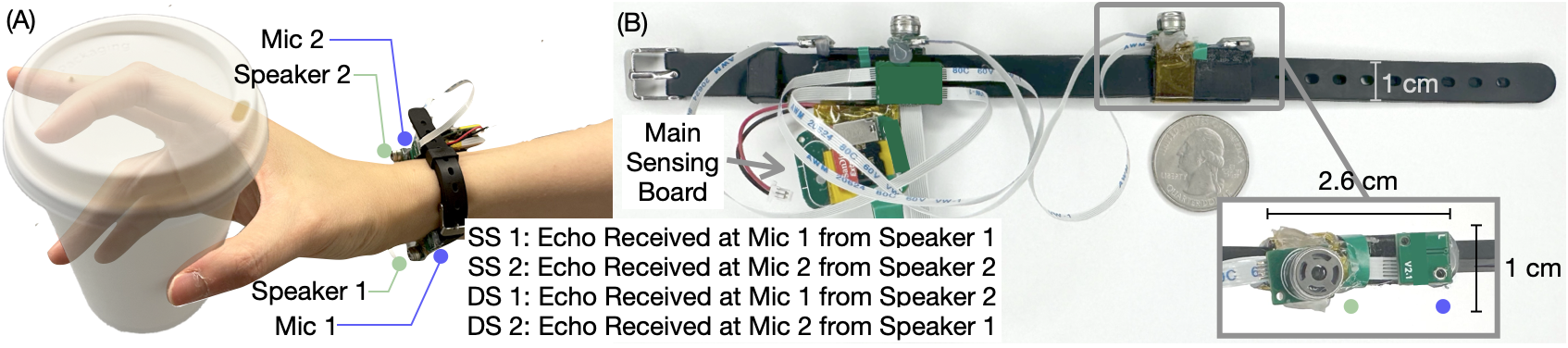}
  \caption{\name{} Prototype.}
  \Description{
  This figure shows the components of the Grab-n-Go System mounted on a wristband. There are two sensing units (highlighted in purple boxes) at each end of the wristband, each containing a microphone and a speaker. The main sensing board is positioned in the middle of the wristband and is secured by slidable holders, allowing the components to be adjusted along the band. A U.S. quarter is placed next to the wristband for size reference, showing that each sensing unit is 2.6 cm long and 1 cm wide. Additionally, there is a 1 cm width reference marked on the wristband to further emphasize the compact size of the system components.
  }
  \label{fig:prototype}
\end{figure*}

\subsection{Hardware Design}    \label{sec:hardware-design}
We designed \name{} to be a small, compact, and low-power device, ensuring its suitability for everyday use. The device is built into a silicone wristband, which is commonly used for watches, offering comfort and flexibility. To accommodate varying wrist sizes, the prototype is designed with an adjustable sensor layout, allowing for easy customization to fit different users.

The system incorporates two speaker-microphone pairs (OWR-06944T-16B and ICS-43434), which are mounted on customized printed circuit boards (PCBs) specifically designed for the sensors. These sensors are strategically positioned to face the hand, enabling them to capture detailed information about both the hand’s movements and the object being held. This arrangement allows the system to detect subtle changes in the acoustic signals reflected by the hand and the object, which are crucial for accurate microgesture recognition. Each speaker-microphone pair is housed in a small 3D-printed case that securely connects the PCBs to the silicone wristband. This case is designed to slide along the wristband, enabling straightforward adjustment of the sensor placement to optimize performance for different users. The sensors are connected to a customized microcontroller module, which includes an SGW1110 module and an MAX98357A audio amplifier. These components are linked via flexible printed circuit (FPC) ribbons.

Powered by a LiPo battery, the system operates by having the microcontroller drive the speakers to emit sound waves while simultaneously collecting the reflected acoustic signals with the integrated microphones. Collected signal data can be stored on a microSD card for offline analysis or transmitted to a smartphone in real time via Bluetooth Low Energy (BLE) for immediate processing.

\begin{figure*}[t]
  \centering
\includegraphics[width=1\textwidth]{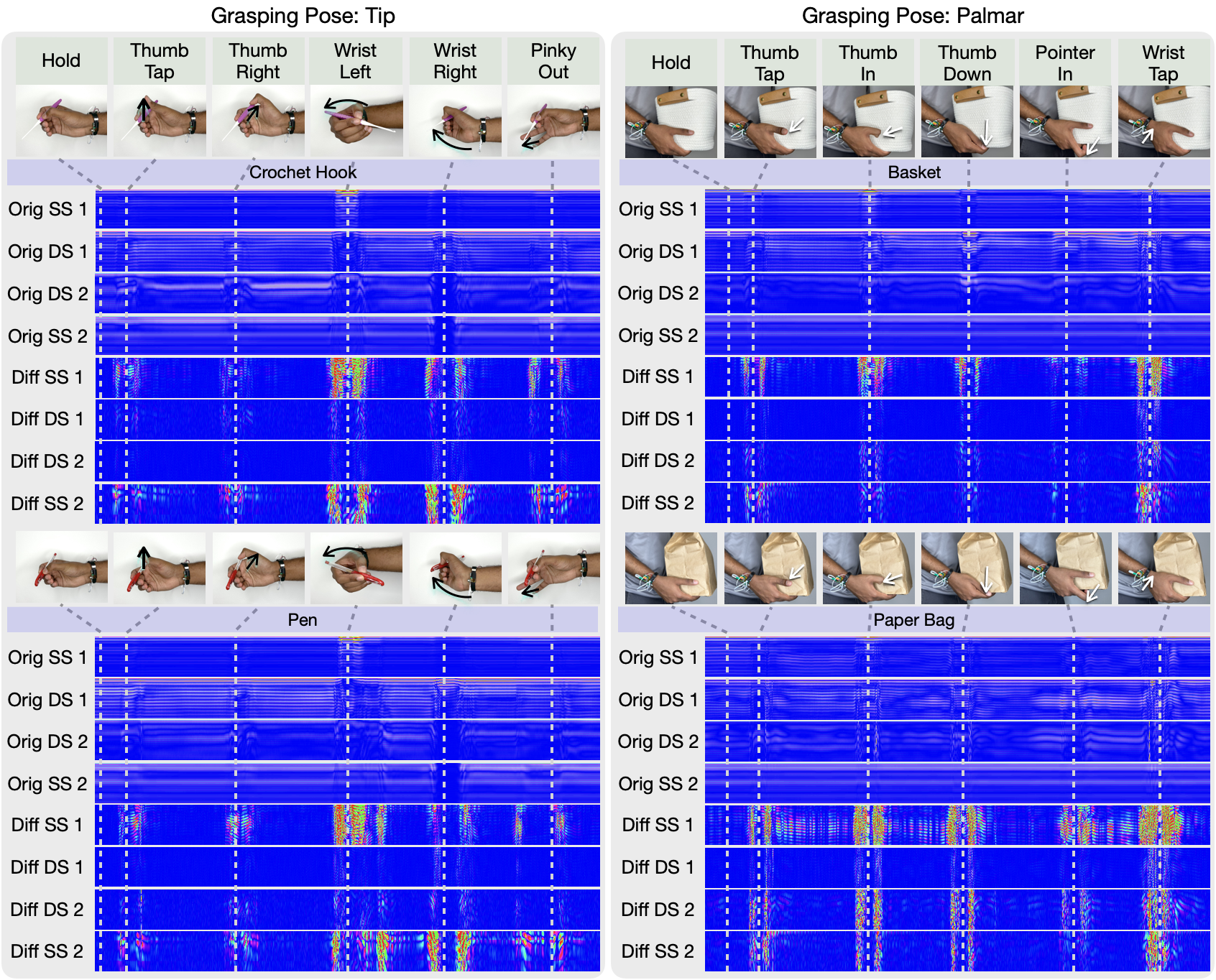}
  \caption{\name{} Example Signals. The microphones capture the acoustic signal emitted from the speaker on the Same Side (SS) and Different Side (DS), resulting in 4 channels (= 2 (microphones) $\times$ 2 (speakers)) of Original (Orig) echo profiles. Subtracting the previous echo profile from the current one creates Differential (Diff) echo profiles, which focus on the movements. In total, 4 channels of original echo profiles and 4 channels of differential echo profiles are used for the system.}
  \Description{This figure shows the echo profiles captured by the active acoustic sensing system of Grab ‘n’ Go. The figure is divided into two sections: the left side of the figure shows the Tip grasp, and the right side shows the Palmar grasp, with each section showing various hand postures used to interact with objects. On the left, the Tip grasp is demonstrated with a Crochet Hook and a Pen, while on the right, the Palmar grasp is demonstrated with a Basket and a Paper Bag. Each grasp type involves six specific hand postures, or micro-gestures. For each grasp type, six micro-gestures are depicted: for Tip, these are “Hold,” “Thumb Tap,” “Thumb Right,” “Wrist Left,” “Wrist Right,” and “Pinky Out”; for Palmar, the micro-gestures are “Hold,” “Thumb Tap,” “Thumb In,” “Thumb Down,” “Pointer In,” and “Wrist Tap.”
  There are vertical dotted lines throughout the figure that separate each hand posture or micro-gesture. These lines help show where each gesture begins and ends. Inside each section, there are eight signal labels that display how sound reflections change during the movement of the hand: Orig SS 1 (Original Same Side 1), Orig DS 1 (Original Different Side 1), Orig DS 2 (Original Different Side 2), Orig SS 2 (Original Same Side 2), Diff SS 1 (Difference Same Side 1), Diff DS 1 (Difference Different Side 1), Diff DS 2 (Difference Different Side 2), and Diff SS 2 (Difference Same Side 2). These labels help explain how the original and changing sound reflections vary on both the side of the wrist facing the object (Same Side) and the side facing away (Different Side).
  Several key patterns can be seen in the echo profiles. The Crochet Hook and Pen have similar profiles due to using the Tip grasp, while the Basket and Paper Bag show similar patterns with the Palmar grasp. For Tip grasps, wrist movements like Wrist Left and Wrist Right produce strong signals in the Diff SS sections because these movements significantly alter hand shape. In Palmar grasps, thumb movements such as Thumb Tap and Thumb In generate strong signals, affecting reflections from both sides of the wrist due to changes in grip and hand shape.}
  \label{fig:example-signals}
\end{figure*}

The two speakers emit acoustic signals of different frequency ranges: 18-21 kHz for one speaker and 21.5-24.5 kHz for the other. This frequency separation ensures that the signals captured by the microphones can be distinctly identified based on their respective frequencies. Specifically, the signals captured by the microphone placed on the Same Side (SS) and Different Side (DS) of the speaker can be differentiated by applying different band-pass filters (Fig. \ref{fig:prototype}). Leveraging these four distinct acoustic signal travel paths --- each corresponding to a different route the sound waves take from the speakers to the microphones --- the system can capture comprehensive information about both the hand and the objects held in it (Fig. \ref{fig:example-signals}). This multi-path signal processing enables a richer, more nuanced understanding of the user’s actions.

\subsection{Theory of Operation}
\name{} is powered by active acoustic sensing. As detailed in the Hardware Design Section (Sec.~\ref{sec:hardware-design}), we placed two speaker-microphone pairs on the wrist, facing the hand. These speakers emit frequency-modulated continuous waves (FMCW) towards the hand and the object being held. The geometry of both the hand-grasping poses and the objects creates a special reflection medium for the acoustic waves, resulting in distinct patterns in the captured signals. As shown in Fig.~\ref{fig:example-signals}, when the user Holds different objects, the captured signals present different characteristics. However, due to the hand's closer proximity to the sensors, the grasping pose introduces a more dominant influence on the captured signal compared to the object itself. Notably, our grasping pose categorization method is based on the overall object geometry, ensuring that objects within the same grasp category tend to produce similar acoustic reflection patterns. This inherent characteristic of the system facilitates the generalization of microgesture recognition across a wide range of objects.

Unlike conventional passive audio analysis, the acoustic signal reflection patterns are formalized using a correlation-based FMCW (C-FMCW) approach called \textit{echo profile analysis}, which is based on the cross-correlation between the transmitted and received acoustic signals \cite{wang2018c}. This technique has proven effective in tracking body part movements when deployed on various wearable devices \cite{li2022eario, mahmud2023posesonic, lee2024echowrist}. In \name{}, each frequency sweep has a duration of 12 ms. We perform cross-correlation between the transmitted signal and the band-pass filtered received signal to extract the signal strengths at different return times. Subsequently, by mapping these time-domain results into the distance domain, using the known speed of sound, we generate the \textit{echo profiles} (Fig. \ref{fig:example-signals}). 

In the echo profiles, each pixel's value represents the correlation strength, which reflects the intensity of the returned acoustic signal. The x-axis of the echo profile corresponds to time, with 12 ms per pixel, while the y-axis represents distance, which is 3.43 mm per pixel. A bright strip in the echo profile indicates a strong reflection at a specific distance. As observed in the example signals (Fig.~\ref{fig:example-signals}), the echo profiles of larger objects, such as the Basket and Paper Bag, present broader and thicker bright strips compared to those of smaller objects like the Crochet Hook and Pen, reflecting the increased surface area for acoustic signal reflections. This approach differs from passive acoustic sensing techniques that typically employ Mel spectrograms to model human auditory perception by reweighting frequency bands according to psychoacoustic principles. Such representations would be suboptimal for our application as they compress precisely the high-frequency information critical for discriminating subtle finger movements. Instead, our echo profiles preserve the spatiotemporal reflection characteristics that directly correspond to physical microgesture execution, capturing the complex interplay between hand configuration and object geometry rather than ambient acoustic events. 

Patterns within the echo profiles reveal changes in the distribution of reflection strengths over different distances and times. To isolate the movements of the hand during microgestures from constant environmental reflections and the static presence of held objects, we calculate \textit{differential echo profiles}. This is achieved by subtracting the preceding pixel value from the current pixel in the echo profile, thereby emphasizing changes over time. As demonstrated in the example signals (Fig.~\ref{fig:example-signals}), the differential echo profiles amplify the changes in echo profiles, which directly correspond to hand movements. For example, when the user is statically Holding the objects, the differential echo profiles present minimal patterns, whereas the patterns are obvious during the execution of dynamic microgestures. Notably, while different hand microgestures and grasping poses yield unique echo profile patterns, objects held in the same grasp pose tend to produce similar features. This characteristic is key to reliably recognizing hand microgestures across a variety of objects.

\subsection{Machine Learning Pipeline}

\subsubsection{Input}
To recognize microgestures, we developed a customized deep-learning pipeline. After the echo profile analysis, these microgestures are represented as different patterns in the \textit{echo profiles}, which lay out as 2D feature maps, as demonstrated in Fig. \ref{fig:example-signals}. By cropping the echo profiles to the window of interest, we obtain input data with a size of 155 $\times$ 70 $\times$ 8. This input tensor comprises 1.8 seconds of temporal data (155 pixels along the time axis), a 24 cm range of interest (70 pixels along the distance axis), and 8 stacked channels: four echo profiles and their corresponding four differential echo profiles.

\subsubsection{Model Architecture}
We propose an Encoder-Decoder model architecture. Given the proven effectiveness of Convolutional Neural Networks (CNNs) in decoding 2D information such as images, we select  ResNet-18~\cite{he2016deep} as the encoder backbone of our model. We incorporate an adaptive 2D average pooling layer with an output size of [1, 1], a dropout layer with a rate of 0.6 to prevent overfitting, and a fully connected layer with an output dimension of 30 to classify the 30 microgestures. Cross-entropy (CE) loss is employed as the optimization objective. The model is configured with an initial learning rate of 0.0002 and a batch size of 8.

\subsubsection{Data Augmentation}
To address potential variations in hand sizes and device positioning --- including user differences and changes after remounting the device --- we incorporate data augmentation techniques into our training process: (a) Vertical Shifting: Echo profiles were randomly shifted vertically by up to 6 pixels to account for slight variations in sensor-to-hand distances. (b) Amplitude Jitter: In 80\% of training iterations, each pixel's intensity value was multiplied by a random factor between 0.95 and 1.05. This amplitude jitter introduces variability in the training data, preventing the model from overfitting to specific signal amplitudes and improving its robustness to noise.

\section{User Study}
To evaluate \name{}'s microgesture recognition performance when holding various objects (Sec.~\ref{sec:object-gesture-set}), we conducted a user study approved by the Institutional Review Board (IRB). 
We recruited 10 participants (3 self-identified as male, 7 as female; age: mean = 24.1, std = 4.04) with a wide variety of hand sizes and shapes (fingertip-to-wrist length: thumb: mean = 126 mm, std = 12 mm; pointer: mean = 170 mm, std = 12 mm; middle: mean = 177 mm, std = 11 mm; ring: mean = 165 mm, std = 12 mm; pinky: mean = 146 mm, std = 10 mm). Note that due to hardware issues, data from 3 of the original 13 participants was broken, resulting in their removal from the study and leading to 10 valid participants. 
Among the participants, one self-identified as ambidextrous, while the remaining nine were right-handed. Since the microgestures are designed to be easily performed with either hand, all participants were instructed to wear the device on their right wrist to maintain consistency in evaluation.
Each study lasted approximately 2 hours, and participants were compensated US\$25 for their time.
The study followed a structured process: participants first completed a demographic survey and hand-size measurements, followed by the primary data collection (Sec.~\ref{sec:data-collection}), and ended with a wearability survey to assess user comfort and experience.

\subsection{Object Set}
To evaluate \name{}'s generalizability across a diverse range of objects, we selected five everyday objects for each grasping pose, as shown in Fig.~\ref{fig:objects-gestures}. These objects exhibit a wide range of shapes, materials, sizes, and weights, reflecting the diversity encountered in real-world scenarios.
To balance study duration while maximizing object diversity, each participant was randomly assigned 2 out of the 5 objects within each grasping pose category. Each object was then tested by four different participants, ensuring balanced data collection. Moreover, we carefully avoided repeated combinations of objects across participants to maintain objectivity and minimize potential biases in the evaluation.

\subsection{Data Collection Procedure}\label{sec:data-collection}
\subsubsection{Apparatus}
The participants stood in front of a laptop (Apple MacBook Pro 14-inch, 2021) placed on a standing desk in a quiet study room.
The laptop served multiple functions: it recorded hand movements using its built-in camera for ground truth label verification, displayed visual stimuli to signal the start of each data collection session, and provided on-screen instructions to guide participants through the process.
All objects used for grasping were placed either on the desk or on the nearby ground, depending on their size, ensuring easy access while maintaining a natural interaction environment.

\subsubsection{Data Collection Sessions}
For each grasping pose in the order of Cylindrical, Hook, Tip, Palmar, and Spherical, each participant completed 1 practice session followed by 6 data collection sessions.
To synchronize \name{} data with the laptop-recorded ground truth, the researcher initiated and concluded each session with a distinctive cue that is both audible and visible (a tumbler tap).
During each session, the participant first performed 4 repetitions (1 repetition for practice sessions) of the 6 microgestures in a randomized order, using one of the assigned objects within the current grasping pose category.
Each microgesture was performed in a 2-second window.
Then, this process was repeated with the second assigned object.
Between each session, the participant removed and re-wore the device under the researcher's guidance.
In total, there were 2 (objects) $\times$ 5 (grasping poses) $\times$ 6 (sessions) $\times$ 6 (microgestures) $\times$ 4 (repetitions) = 1440 microgesture instances collected from each participant, resulting in a total of 14,400 microgesture instances across all participants. Following data review, 21 instances were relabeled and 45 instances were excluded due to incorrect microgesture execution.

\subsection{Training Scheme}
\subsubsection{Two-Step Training Scheme}    \label{sec:two-step-training-scheme}

\begin{figure*}[ht]
  \centering
  \includegraphics[width=1\textwidth]{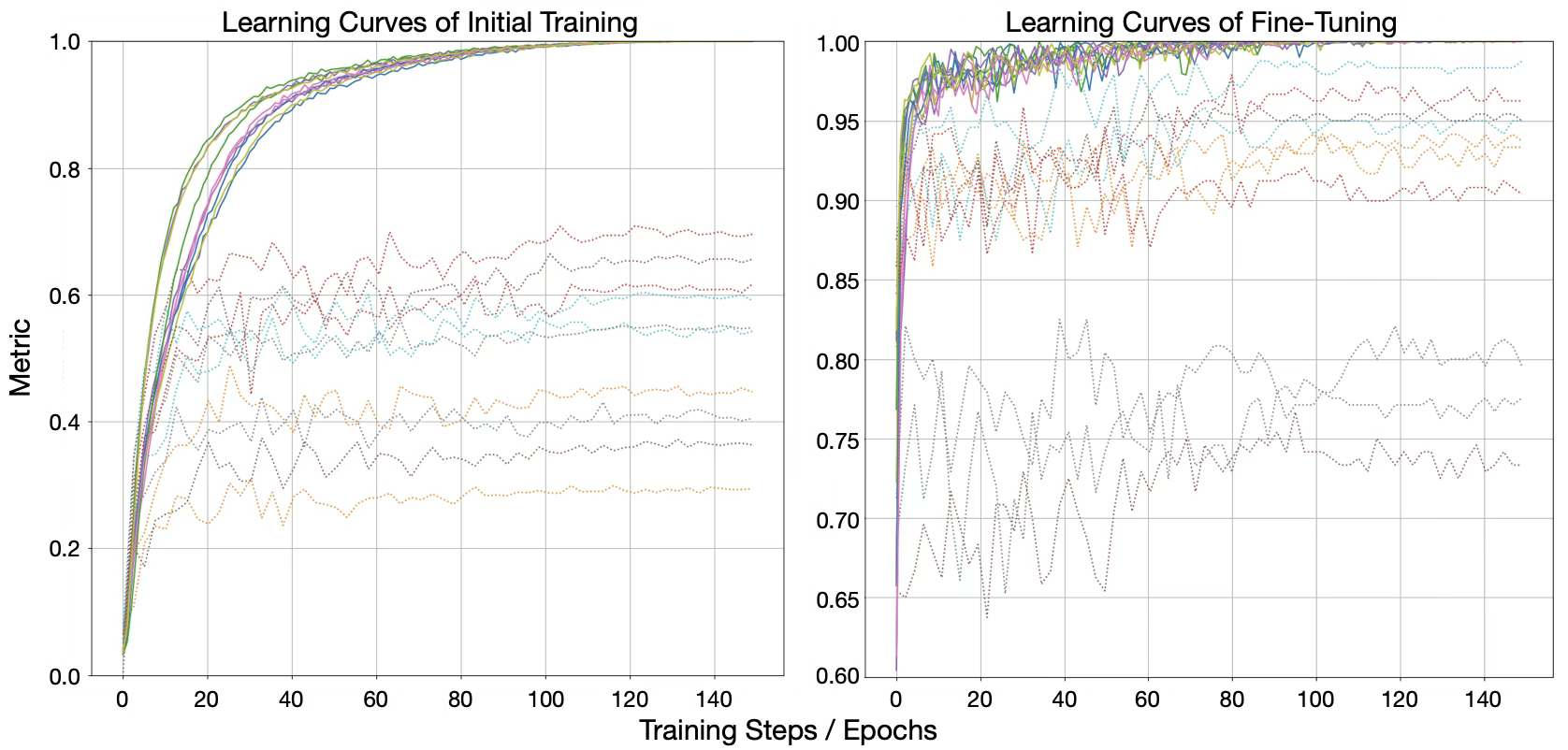}
  \caption{Learning curves for the models trained on user study data. Learning curves for models trained on user study data. Each color represents a distinct participant's model, with solid lines indicating training performance and dotted lines showing testing performance.}
  \Description{
  
  }
  \label{fig:learning-curves}
\end{figure*}

To minimize training efforts for new users and enhance system performance, we implement a two-step training scheme. Although the system remains user-dependent, we optimize efficiency by fine-tuning a pre-trained user-independent model rather than training a customized model from scratch for each user.

Specifically, for each participant in the user study, we first trained a leave-one-participant-out model. This model was trained on data from all other participants, excluding the target participant, and the model was tested on the data from the excluded target participant. This phase lasts 150 epochs and produces a user-independent model that captures generalizable features of microgesture execution. In the second step, we fine-tuned this user-independent model on the target participant's data for an additional 150 epochs, adapting the model to the specific characteristics of the individual user. The choice of 150 epochs for both the initial training and fine-tuning phases was determined through empirical validation during our pilot study, where we observed that model performance typically plateaued with minimal fluctuation around this value for both training and testing sets (Fig.~\ref{fig:learning-curves}). While we did monitor validation performance, we did not formally implement early stopping based on a specific performance increase threshold.

\subsubsection{Wearing Session Independence}    \label{sec:wearing-session-independence}
While not entirely user-independent, \name{} supports wearing session independence. This crucial feature eliminates the need for repeated data collection and model retraining each time the user removes and re-wears the device. This is particularly important in real-world scenarios where users will inevitably remove the device for charging or other reasons. With wearing session independence, users only need to provide a one-time data collection upon initial device acquisition, mirroring the familiar process of finger ID or face ID registration.

To evaluate wearing session independence, each participant in the user study collected data across six sessions for each grasping pose. Between each session, the participant removed and re-wore the device, simulating real-world usage scenarios. During the training phase, the model was trained on data from five of the six sessions and subsequently tested with the remaining session. To mitigate the potential influence of user familiarity with the microgestures over time, the final performance metric was calculated by averaging the results obtained from all possible combinations of training and testing sessions.

\subsection{Results}

\begin{figure*}[ht]
  \centering
  \includegraphics[width=1\textwidth]{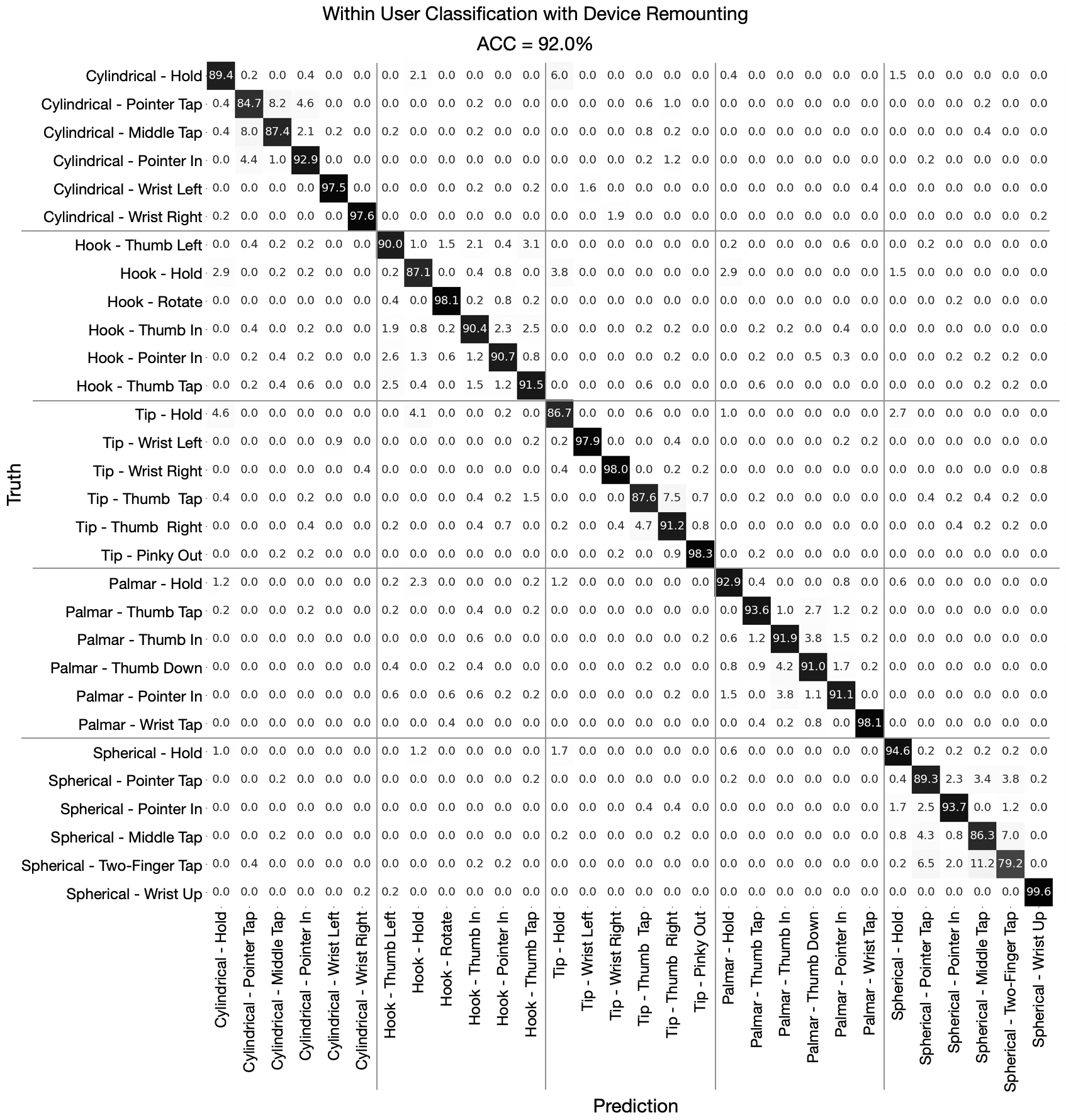}
  \caption{The confusion matrix of the study results.}
  \Description{
  This figure presents a confusion matrix that displays the true gestures on the vertical axis (labeled “Truth”) and the predicted gestures on the horizontal axis (labeled “Prediction”). Each square in the matrix represents the frequency of a combination of true and predicted gestures. The color intensity of each square indicates the prediction accuracy, with darker squares reflecting higher accuracy.
  The diagonal entries, running from the top-left to the bottom-right, represent instances where the system correctly predicted the hand gestures. These entries are noticeably darker, signifying a high level of accuracy in predicting the corresponding gestures.
  In contrast, the off-diagonal entries represent misclassifications, where the system’s predicted gesture did not match the actual gesture. These entries are lighter in color, indicating lower prediction accuracy for those specific gesture pairs.
  Overall, the confusion matrix demonstrates an accuracy of 92.0\% (as indicated at the top), suggesting that the system performs well in recognizing most hand gestures with a high degree of reliability.
  }
  \label{fig:cm-all-fine-tuned}
\end{figure*}


\subsubsection{Training Scheme}
To ensure the study remained manageable for the participants, we limited the number of microgesture instances collected per participant to 24 per microgesture per object (= 4 (repetitions) $\times$ 6 (sessions)), with the device being remounted between sessions. Given the relatively small size of this dataset, we employed a two-step fine-tuning training scheme (Sec.~\ref{sec:two-step-training-scheme}) to assess the potential benefits of larger base datasets. Initially, a leave-one-participant-out (LOPO) model was trained for each participant using data from the remaining 9 participants. Subsequently, to evaluate the system's wearing session independence, this LOPO model was fine-tuned using the leave-one-session-out (LOSO) method (Sec.~\ref{sec:wearing-session-independence}) for each participant. In each iteration, the LOPO model was fine-tuned using data from 5 sessions and tested on the held-out session for each grasping pose. This process was repeated 6 times, with each session serving as the held-out set in turn. 

\subsubsection{Microgesture Recognition Results}
Overall, by averaging the results across all the participants and wearing sessions, the fine-tuning training process achieved an average accuracy of 92.0\% in recognizing 30 microgestures performed on 25 different objects (Fig. \ref{fig:cm-all-fine-tuned}). Importantly, each participant interacted with a unique and randomly assigned combination of objects for each grasping pose, ensuring no two participants encountered the exact same object set. In addition, our object set included a wide variety of shapes, materials, sizes, and weights, reflecting the complexity of real-world interactions. The system’s ability to maintain high accuracy across 25 different objects highlights its strong cross-object generalizability.

Beyond the encouraging average accuracy of our system, the results also demonstrate a low average false-positive rate of 0.2\%. This metric, calculated as the ratio of False Positives to the sum of False Positives and True Negatives, is particularly critical for the practical usability of microgesture-based interaction systems. False activations represent one of the most substantial barriers to user acceptance and system reliability in real-world contexts. A low false-positive rate signifies that the likelihood of the system erroneously detecting a microgesture when none was intended is minimal. This is important for reducing user frustration and fostering a reliable user experience. The low false-positive rate further underscores the feasibility of \name{} for real-world deployment, as it suggests a minimal tendency to trigger unintended actions, even when interacting with a diverse array of everyday objects. This reliability is essential for users to integrate microgestural input into their daily routines without concerns about spurious activations. However, we also want to admit that this low false-positive rate was achieved in a relatively controlled lab environment. Further experiments and studies will be needed to test the system in real-world scenarios where the false-positive rate can likely be higher due to the huge variance of daily body postures and noise in real-world settings.


\begin{figure*}[ht]
  \centering
  \includegraphics[width=1\textwidth]{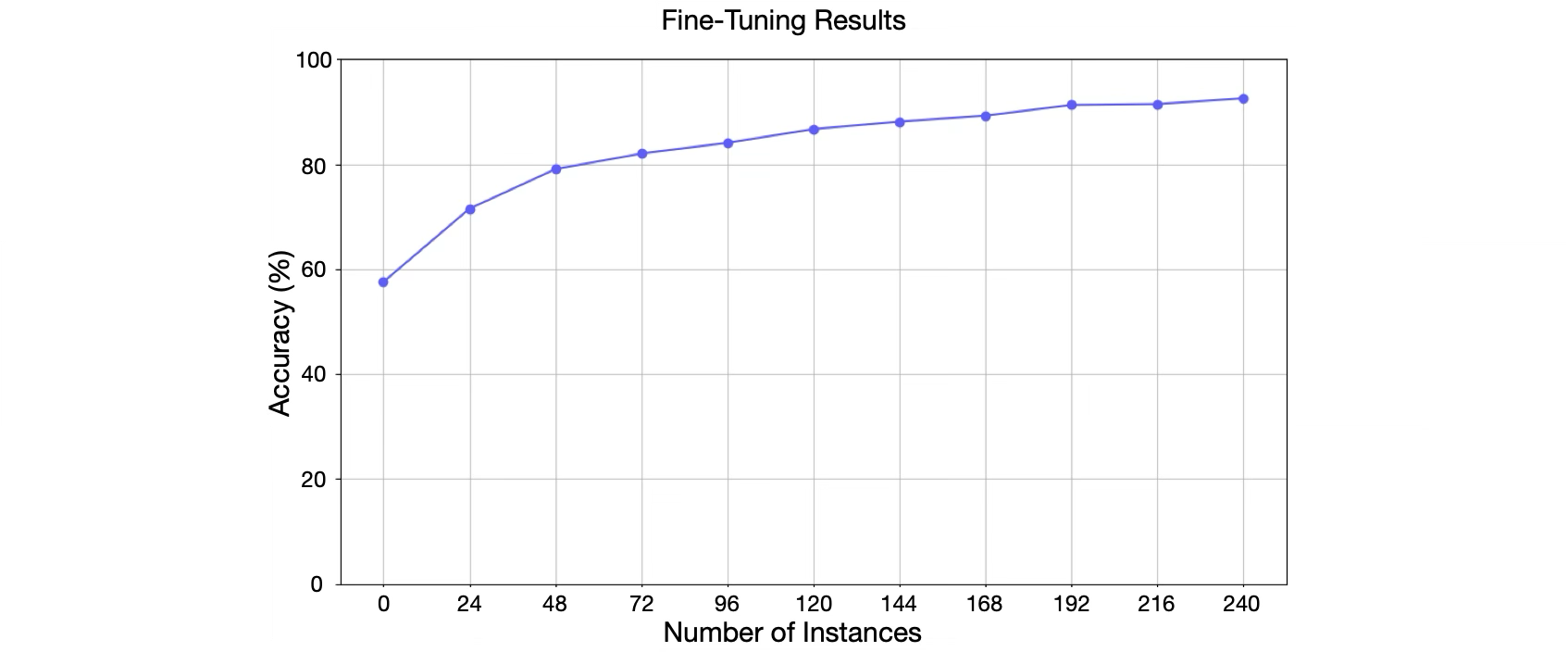}
  \caption{The performance when fine-tuning the model with different amounts of data.}
  \Description{
  
  }
  \label{fig:fine-tune}
\end{figure*}

\subsubsection{Fine-Tuning Results}
To assess the impact of our two-step fine-tuning training scheme, we evaluated the system's performance by directly training user-dependent models using only the data from each of the 10 individual participants. This purely within-user training involved using 5 sessions of data for training and the remaining 1 session for testing, all belonging to the same participant. Averaging the results across all participants and wearing sessions, this direct training approach yielded an average accuracy of 86.19\%, a 5.81\% decrease compared to the fine-tuning results (92.0\%). This difference underscores the benefit of leveraging the foundation model trained on a broader dataset. Furthermore, it suggests that with the acquisition of even larger and more diverse datasets in the future, there is considerable potential for further performance improvements.

Since collecting personalized training data can be time-consuming and inconvenient, reducing the amount of required data is crucial for improving user adoption. To minimize the amount of training data required from new users, we investigated how much data is necessary to fine-tune the model without significantly compromising performance. We systematically varied the number of sessions, or microgesture instances, used during the fine-tuning process to assess its impact on recognition accuracy.

As shown in Fig~\ref{fig:fine-tune}, with a fully user-independent model, where no personalized data from the new user was used, the average accuracy was 57.5\%, indicating that while the model captured some generalizable features, it struggled to adapt to individual differences. However, when incorporating just one session of data (24 microgesture instances per grasping pose), average accuracy increased to 71.4\%, demonstrating that even a small amount of user-specific data enhances performance. The model exceeded 80\% accuracy with three sessions (72 instances) and surpassed 90\% accuracy with eight sessions (192 instances).

These results suggest a promising potential to reduce the required fine-tuning data. We believe that as the base user-independent model is trained with more data, the amount of data needed from new users can be further reduced, making \name{} more practical for real-world deployment.

\subsubsection{Object-Independent Results}  \label{sec:object-independent-results}
To further investigate the generalizability of \name{} across various objects, we conducted an object-independent evaluation. Within each grasping pose category, we trained the model using data from four objects and tested it on the remaining object. This process was repeated for all the objects. 

The average accuracy across the 25 objects is 85.3\% (Cylindrical: 85.1\%, Hook: 81.0\%, Tip: 93.7\%, Palmar: 87.1\%, and Spherical: 79.5\%). It is important to note that each of these object-independent models was trained on a relatively limited dataset of 576 microgesture instances (= 6 (microgestures) $\times$ 4 (repetitions) $\times$ 6 (sessions) $\times$ 4 (participants)) are used for training. In addition, the data came from different participants, introducing more variability and potential uncertainty, which could affect the performance. Given these factors, we believe that further improvements could be achieved by incorporating more data from the same participant or by including additional objects within the same grasping pose. This would help the model learn a richer set of features, ultimately boosting its performance and robustness across various objects and users.

\subsubsection{Discussion}
Despite the use of similar microgestures for different grasping poses, e.g., Pointer Tap was used for Cylindrical and Spherical, and Pointer In was used for Cylindrical, Spherical, Palmar, and Hook, \name{} successfully differentiated between them. This demonstrates the system's ability to recognize not only hand movements but also grasping poses. Analyzing the confusion matrix, incorrect predictions primarily occurred within the same grasping pose category. This indicates that \name{} effectively classified distinct grasping poses, suggesting potential for further refinement by fine-tuning the model on specific grasping pose categories to enhance performance.

Our findings indicate that wrist-related microgestures exhibit the highest recognition accuracy. When examining the echo profiles (Fig.~\ref{fig:example-signals}), these microgestures produce the most distinguishable signals, likely due to their proximity to the sensors and relatively large scale of movements on the palm. Despite being farther from the sensors and generating weaker signals, finger-related microgestures still achieve satisfactory performance. This suggests the potential for incorporating additional finger-based microgestures in future iterations.

Pointer Tap and Middle Tap exhibited the highest levels of confusion, which is unsurprising given the proximity of these two fingers and the similarity of their movements. In addition, between each session, the participant not only removed and re-wore the device but also re-held the object, introducing additional sources of variation, which was designed to simulate real-world scenarios. Despite these challenges, the majority of recognition accuracies remained above 92\%. When analyzing performance within each grasping pose, Tip exhibited the least confusion, while Spherical demonstrated the most. This observation aligns with the observation that microgestures performed closer to the sensors can yield better accuracy. This insight provides valuable guidance for designing future wrist-worn active acoustic sensing devices and microgesture sets.

Overall, according to our user study evaluation results, \name{} effectively recognizes 30 microgestures across 25 distinct everyday objects using only a single wristband, positively supporting our proposed research question as described earlier in the paper.

\subsection{Follow-Up Study with Deformable Objects}

\begin{figure*}[t]
  \centering
  \includegraphics[width=1\textwidth]{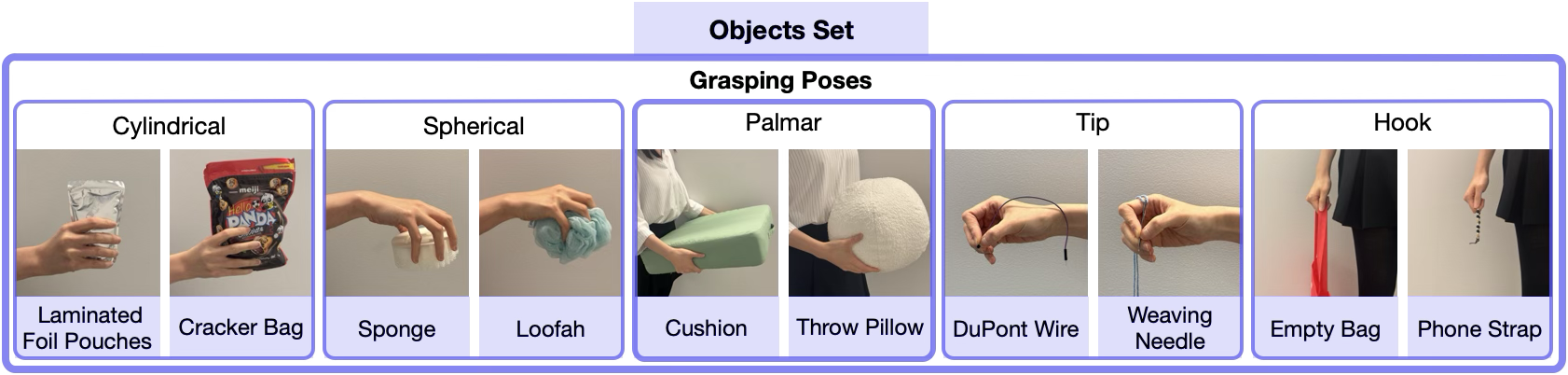}
  \caption{The object set of the follow-up study.}
  \Description{
  
  }
  \label{fig:study-2-objects}
\end{figure*}

While \name{} effectively captures object and hand shape for microgesture recognition, deformable objects present a challenge to sensing accuracy. Unlike rigid objects that maintain consistent acoustic reflection patterns, deformable materials introduce variability through shape distortions that occur during natural manipulation. Consequently, we conducted a follow-up study to assess the impact of these shape variations on performance.

This study involved 8 right-handed participants (3 self-identified as male, 5 as female; age: mean = 24.25, std = 3.45; fingertip-to-wrist length: thumb: mean = 131 mm, std = 8 mm; pointer: mean = 171 mm, std = 7 mm; middle: mean = 179 mm, std = 9 mm; ring: mean = 167 mm, std = 8 mm; pinky: mean = 144 mm, std = 9 mm). Consistent with our initial protocol, participants wore the prototype device on their right wrist throughout the study. Each study lasted approximately 1.5 hours, and participants received compensation of US\$25 for their participation.

\subsubsection{Study Setup} Maintaining the same process and apparatus as the initial study, we introduced a different object set comprising two deformable everyday items for each grasping pose (Fig.~\ref{fig:study-2-objects}). These objects were specifically chosen due to two key challenges: (1) their shape varied with each grasp, and (2) they deformed further during microgesture execution. Each participant interacted with one object per grasping pose category, and each object was evaluated by four distinct participants, aligning with the initial study's protocol. 

\subsubsection{Data Collection Procedure}
This study mirrored the procedure of the initial study, with the key modification that participants used a single object for all sessions within a specific grasping pose category. Consequently, each of the eight participants contributed 720 microgesture instances (= 1 (objects) $\times$ 5 (grasping poses) $\times$ 6 (sessions) $\times$ 6 (microgestures) $\times$ 4 (repetitions)), resulting in a dataset of 5,760 microgesture instances across all eight participants. Following data review, 30 instances were relabeled and 38 instances were excluded due to incorrect microgesture execution.

\subsubsection{Results}

\begin{figure*}[t]
  \centering
  \includegraphics[width=1\textwidth]{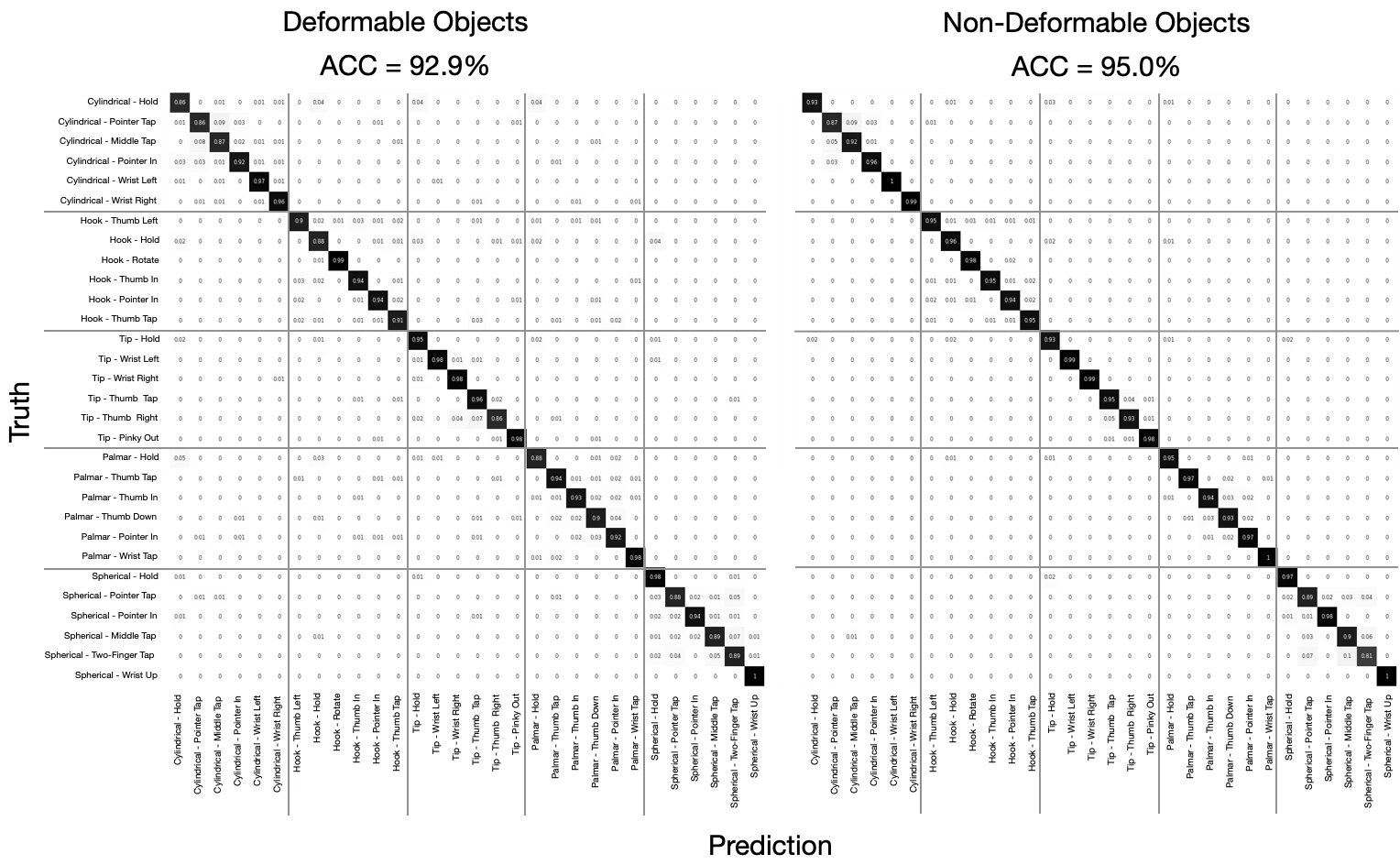}
  \caption{The confusion matrix of the study results of the follow-up study.}
  \Description{
  
  }
  \label{fig:cm-study2}
\end{figure*}

Using the data from the initial study and the follow-up study, we created a joint dataset incorporating data from all 18 participants. Employing our two-step fine-tuning training scheme (Sec.~\ref{sec:two-step-training-scheme}), we initially trained 18 leave-one-participant-out (LOPO) models on this combined dataset. Note that as the follow-up study tested only a single object per grasping pose per participant, the fine-tuning phase for participants from the initial study utilized only the data corresponding to the first object they were tested with for each grasping pose. These user-independent models were subsequently fine-tuned using individual participant data containing one object per grasping pose. 

Our system achieved an average microgesture recognition accuracy of 95.0\% with non-deformable objects, which slightly decreased to 92.9\% when interacting with deformable objects. The corresponding average false-positive rates were 0.2\% and 0.3\%, respectively. Note that the better performance observed with non-deformable objects compared to the initial study can be attributed to two factors: (1) the base model benefits from being trained on a larger, combined dataset from both studies; (2) the evaluation utilized only one object per grasping pose, in contrast to the two objects used in the initial study. For the deformable objects, the lowest individual object accuracy was observed with the laminated foil pouches, primarily attributed to their inherent softness and significant shape variation along the grasping area. The pouches exhibited different deformation patterns with each grasp, and variations in grasping height further contributed to inconsistencies across data collection sessions. The second-lowest accuracy was recorded with the loofah, likely due to its extreme softness leading to high deformation variability during microgesture execution. Additionally, its slippery surface likely contributed to gradual slippage during data collection, particularly for participants with smaller hands. We acknowledge this inherent challenge posed by deformable objects; however, given the relatively small overall performance difference, we remain optimistic that the user experience in real-world applications will not be significantly impacted.

To facilitate further research in this domain, we released the whole dataset with 18 participants and 35 objects to the research community.

\subsection{Post-Study Wearability Survey Results}

Given the identical apparatus used in both studies, we analyzed the wearing experience collectively. Firstly, none of the participants reported perceiving any sound from the device. This indicates that our frequency sweep is compatible with commercial speakers and microphones while remaining inaudible to users. Secondly, the participants generally found \name{} comfortable to wear (mean = 5.89, std = 1.18 on the Likert scale: 1 = extremely uncomfortable, 7 = extremely comfortable). Specifically, several participants described it as comfortable (P1-3, P1-4, P1-5, P1-12, P2-1, P2-2), fitting well (P1-7, P2-3, P2-7), and feeling like a normal wristband or watch (P1-1, P1-13, P2-7). However, one participant noted that the 3D-printed component felt slightly sharp (P2-8), and another experienced mild skin irritation from the rubber strap (P1-11). For future iterations and deployment, careful attention should be paid to the device's edges to eliminate any sharpness, and alternative strap materials should be explored to enhance user comfort for a wider range of skin sensitivities. 

\section{Discussion}
The performance of \name{} in the user study is promising as a proof of concept. The proposed methods can be further optimized for real-world applications, and we aim to discuss some of the key points for future improvements.

\subsection{Hands-Busy Interaction}
With the increasing ubiquity of computing devices, traditional input modalities that rely heavily on explicit manual dexterity create conflicts between digital interaction and everyday physical tasks. This tension forces users to continually prioritize between their current hands-on activities and device manipulation. To address this challenge and support always-available input, researchers have investigated various microgestural interaction paradigms that can be seamlessly integrated into everyday scenarios where hands are already occupied \cite{sharma2019grasping, sharma2021solofinger, sharma2024graspui}. Building upon these prior microgesture design strategies, we defined our own microgesture set for evaluation. 

Recognizing the importance of hands-busy interaction, specifically input modalities that allow interaction even when hands are occupied, researchers have investigated various sensing modalities for microgesture recognition in such situations. Saponas et al.~\cite{saponas2009enabling} utilized forearm EMG on an armband to recognize the pressing of four distinct fingers while holding a travel mug or bag. Rudolph et al.~\cite{rudolph2022sensing} proposed a capacitive sensing wristband capable of distinguishing dynamic object interactions like slide, rock, twiddle, squeeze, stretch, and tripod pinch, with each interaction tied to a specific kind of object. VibAware~\cite{kim2023vibaware} employed bio-acoustic sensing with a ring-wristband combination to detect thumb and index finger taps and swipes when grasping four specific 3D-printed props. SparseIMU~\cite{sharma2023sparseimu} used distributed IMUs across finger joints to recognize six different microgestures while holding 12 distinct objects. However, the presence of objects introduces significant challenges, including diverse object geometries and potential occlusion, often leading to limitations in the variety of objects and microgestures investigated. While these prior works presented promising results within their specific contexts, the question of how well microgesture recognition with busy hands can generalize across a wide and diverse range of objects remains largely unanswered. This formed our primary design consideration (Sec.~\ref{sec:object-gesture-set}). By testing \name{} with 35 objects, encompassing both solid and deformable types, we verified the generalizability of \name{}, supporting its potential for future deployment in everyday life where users interact with a multitude of items. In addition, tackling our second design consideration regarding sensing technique and form factor (Sec.~\ref{sec:choice-of-technique}), \name{} achieves a promising performance while maintaining the lightweight watch-like form-factor. This preserves the potential integration with commercial smartwatches in the future. The comparison with prior work is summarized in Table~\ref{tab:comparison_table}. 

\subsection{Usage Scenarios}
As discussed in the previous section, enabling hands-busy interaction across a variety of everyday objects is a crucial aspect of ubiquitous computing. With \name{}, we envisioned that the users could interact with their computers with subtle hand gestures even when their hands are occupied. This applies to short, fast, and discreet interaction scenarios. To illustrate a potential use case, consider Lisa, who is rushing to work and needs a coffee to stay alert. With her handbag in her left hand and a freshly purchased coffee in her right, she desires to switch her credit card for payment. Rather than struggling to find a place to set things down and swipe on her smartwatch, she can elegantly switch the credit card using the Pointer In microgesture on the coffee cup (Fig. \ref{fig:teaser}).

In another scenario, Lisa is inspecting newly soldered prototypes and keeping relevant records on her laptop. To efficiently mark each soldering point after testing, she can use Wrist Right and Wrist Left microgestures to navigate between recording items. Thumb Tap can be used to mark a point as connected, while Pincky Out can indicate a disconnected point. When seeking a break, Lisa can grab her coffee and initiate music playback with a Pointer Tap microgesture, and use Wrist Left and Wrist Right to navigate between songs.

\subsection{Hardware Design}
To effectively capture signals surrounding the hand, we strategically placed two pairs of sensors on each side of the wrist. However, given that certain microgestures, such as those associated with the Spherical grasping pose, occur primarily on one side of the hand, and the opposite-side sensor can be occluded by in-hand objects, it is worth exploring the feasibility of using only a single speaker-microphone pair in future implementations. This approach could potentially reduce the system's complexity and improve its wearability, particularly for applications focused on a specific subset of microgestures.

\subsection{Machine Learning Algorithm Comparison}

Beyond the proposed ResNet-18 architecture, we sought to understand the efficacy of various machine learning algorithms when applied to our C-FMCW-based active acoustic data. We evaluated the performance of commonly used machine learning algorithms on the collected initial user study data (Sec.~\ref{sec:data-collection}). For each model, we employed a consistent training and validation strategy: training on the first five data collection sessions and validating on the final session for each participant. The reported results represent the average performance across all participants. 

Inspired by prior work with a similar goal \cite{saponas2009enabling, rudolph2022sensing, kim2023vibaware, sharma2023sparseimu}, we first evaluated with traditional machine learning methods from the Scikit-learn library \cite{pedregosa2011scikit}. We compared the performance of Linear Support Vector Classification (LinearSVC),  Linear Discriminant Analysis (LDA), and Random Forest. Given our 2D image-like echo profiles, we initially used a flattened version of the echo profiles as input. As an alternative, we explored feature extraction. A pilot study investigated several common image feature extraction techniques, including color histograms, Haralick features, Local Binary Patterns (LBP), and Histogram of Oriented Gradients (HOG). Based on the pilot study, Haralick features yielded the most promising results, and we subsequently used this method to generate an alternative input representation. Overall, training the LDA model with Haralick feature input produces the best results.

\begin{table*}[t]
    \centering
    \caption{Comparison of Different Machine Learning Algorithms}
    \resizebox{\textwidth}{!}{
        \begin{tabular}{|c|c|c|c|c|c|c|c|c|c|c|c|c|c|c|c|}
            \hline
             \textbf{LinearSVC} & \textbf{LinearSVC} & \textbf{LDA} & \textbf{LDA} & \textbf{Random Forest} & \textbf{Random Forest} & \textbf{TabPFN} & \textbf{CNN-LSTM} & \textbf{RepViT} & \textbf{FastViT} & \textbf{ResNet-18} \\
             (Flatten) & (Haralick) & (Flatten) & (Haralick) & (Flatten) & (Haralick) & (Haralick) &  &  &  & \\
            \hline
             45.87\% & 72.33\% & 47.37\% & 73.54\% & 60.32\% & 71.67\% & 78.56\% & 85.60\% & 73.20\% & 85.11\% &  \textbf{86.63\%}\\
            \hline
        \end{tabular}
        }
    \label{tab:ml_table}
\end{table*}

Subsequently, we explored deep learning methods. Given the relatively small size of our custom-collected dataset, we employed TabPFN~\cite{hollmann2022tabpfn}, a pre-trained Transformer specifically designed for supervised classification on small tabular datasets. Leveraging its foundation model, which was trained on a large and diverse corpus of tabular data, our TabPFN model achieved an average accuracy of 78.56\%.

With a focus on potential on-device deployment, we explored lightweight methods. Specifically, we adopted Reparameterized Vision Transformer (RepViT)~\cite{wang2024repvit}, which combines CNN efficiency with Vision Transformer design principles, leveraging its ability to maintain high accuracy while significantly reducing computational demands. We also assessed FastViT~\cite{vasu2023fastvit}, a hybrid architecture that strategically balances CNN and Transformer components to optimize both performance and processing speed. Our experiments demonstrated that RepViT achieved accuracy comparable to traditional methods, while FastViT surpassed them with an accuracy of 85.11\%.

To better leverage the temporal dynamics inherent in the microgesture echo profiles, we developed a customized deep-learning network incorporating a CNN-LSTM encoder augmented with attention mechanisms and a fully connected classifier. This architecture yielded performance comparable to our ResNet-18 model, suggesting the temporal information captured by the LSTM provides a complementary representation of the microgestures.

Our evaluation revealed that our proposed  Encoder-Decoder architecture leveraging a ResNet-18 backbone yielded the best results. However, it is important to highlight that the CNN-LSTM and FastViT achieved results comparable to those of this top-performing configuration. Given FastViT's significantly more lightweight nature in terms of computational complexity and memory footprint compared to the ResNet-18 model, it emerges as a highly promising candidate for future deployment scenarios, particularly on resource-constrained platforms such as mobile devices or even direct on-chip deployment.

\subsection{Overfitting}
In our evaluation, we observed a difference between the final training accuracy, which reached 100.0\%, and the final testing accuracy of 92.0\%. This 8.0\% gap suggests a degree of overfitting, a phenomenon not unexpected given the inherent constraints of our dataset size and the complexity of the 30-class microgesture recognition task across 25 diverse objects. While this level of overfitting warrants consideration for future work, we believe the achieved 92.0\% testing accuracy represents a strong level of real-world performance, potentially meeting the practical threshold for reliable and usable interaction within our intended application context. Nevertheless, we acknowledge that further research into mitigating this overfitting and enhancing the model's generalization capabilities could lead to even more robust and dependable performance in broader deployment scenarios.

\subsection{Real-World Application}
While this paper focuses on microgesture recognition with objects in hand, it is important to acknowledge that hands are not always occupied, and the system should not always be active. We plan to investigate methods for activating microgesture recognition only when necessary. One potential approach involves integrating a separate system to detect hand occupancy. Alternatively, introducing a unique activation gesture that is less likely to occur accidentally in daily life could also be considered. 

As our evaluation was conducted in a controlled in-lab setting, the robustness of the microgesture sets in real-world environments remains to be explored. Some microgestures may be commonly used for other purposes, potentially leading to the accidental activation of unintended functions in daily life. Future research should investigate the system's robustness, following the approach outlined in SoloFinger \cite{sharma2021solofinger}, to ensure its suitability for real-world deployment.

Considering the two-step fine-tuning scheme's implications for usability in future deployments, we envision leveraging a large, shared base dataset for the initial training of a foundational model. This approach would streamline the user experience, requiring new users to collect only a single set of personalized data during their initial device setup, akin to the familiar process of registering face ID on smartphones or configuring new gestural inputs on augmented/virtual reality devices. This minimal initial effort would offer a streamlined and user-friendly onboarding process.

\subsection{The Impact of Object Selections}
While we have demonstrated that \name{} successfully recognizes microgestures across 35 different objects as a proof of concept for a research prototype, the range of objects people interact with every day goes far beyond this number. We plan to further investigate the system's ability to recognize microgestures on unseen objects that share similar grasping poses. This will include assessing whether the current model can accurately classify microgestures on objects not included in our existing dataset. While we presented the object-independent results (Sec.~\ref{sec:object-independent-results}), further exploration with more diverse data within the same grasping pose category is needed. Ultimately, our goal is to enable \name{} to support microgesture recognition across any arbitrary object.

Since we leverage ultrasonic range for our active acoustic sensing method, external noises generally do not interfere with the received acoustic signals. However, when certain materials are taped or scratched, they can produce high-frequency signals. By incorporating a diverse range of materials of objects into our dataset, we effectively mitigated the impact of these signals.

\subsection{Microgesture Set}
Despite providing participants with two seconds to perform each microgesture, the execution speed varied significantly. Additionally, the level of exaggeration in microgestures also differed. However, training the base model on data from all participants effectively addressed these challenges, demonstrating \name{}'s ability to handle diverse microgesture styles.

Given the variation in participant hand sizes and object dimensions, the manner in which objects were held also differed significantly. For instance, while a typical grasping pose for Spherical objects involves an arched palm, a participant with smaller hands found it challenging to grasp the jar lid, leading to a flatter palm. Despite these variations, \name{} effectively adapted to these diverse holding styles.

Although \name{} successfully recognized 30 microgestures, the specific number required may vary depending on the application. For instance, 3 microgestures per grasping pose might be sufficient for quick tasks on commercial smartwatches, like answering calls or muting the device. Moreover, focusing on microgestures associated with a single grasping pose, such as pen interaction with the Tip microgesture when using an Apple Pencil, could be a potential use case. By reducing the microgesture set, we anticipate further improvements in performance.

Our evaluation showcased \name{}'s capability to support a comprehensive set of 30 distinct microgestures. While the design of this set was informed by prior research and considerations for ease of execution across various object shapes and weights, we recognize that such a large set could potentially impose a significant cognitive load on users for memorization in practical applications. The primary goal of our evaluation was to rigorously investigate the fundamental recognition capabilities of \name{} across a wide range of objects, hence our decision to test this extensive microgesture set as a proof of concept. For real-world deployment scenarios, we plan to collaborate closely with user experience researchers to carefully curate a more streamlined and intuitive microgesture set that balances functionality with ease of memorization and use. 

\subsection{Limitation}
While our evaluation was conducted in a controlled lab setting, some participants exhibited movement due to the extended study duration. However, the overall movement is limited. We plan to explore \name{}'s performance during more dynamic activities, such as walking, to gain a deeper understanding of its capabilities in real-world scenarios. 

Additionally, we observed that certain objects, when held in the hand, can significantly obstruct acoustic signals, hindering the recognition of finger microgestures. For instance, when holding a pillow using the Palmar grasping pose, the softness of the pillow can cause the hand, including the sensors, to sink into the material, preventing the propagation of in-air acoustic signals. As a result, the corresponding echo profiles lacked distinct patterns, making microgesture detection unreliable. \name{} may not work well on these objects, which is another limitation of our proposed system. 

In terms of the occlusion, \name{}'s capability will also be constrained by the covering of the clothes. It is natural that sometimes the watch will be covered by the sleeve. However, if being covered by the sleeve, \name{} will suffer from the signals being blocked by the sleeve and can not capture the information from the desired area.
\section{Conclusion}
In this paper, we introduce \name{}, a wristband that enables robust recognition of hand microgestures when hands are occupied with objects. Leveraging active acoustic sensing, \name{} effectively identifies 30 microgestures with an average accuracy of 92\% across a diverse set of 25 objects. The system was evaluated through a user study with 10 participants, involving the collection of 14,400 microgesture instances. A follow-up study with an additional 8 participants further expanded our dataset by collecting 5,760 more microgesture instances, specifically to validate the system's robustness against more challenging, deformable objects. Overall, \name{} showcases its effectiveness in enabling microgesture recognition across a wide range of everyday objects, providing a seamless solution for always-available input.

\begin{acks}
This project was supported by National Science Foundation Grant No. 2239569. We want to thank our colleagues at SciFi Lab for their invaluable support, all participants for their generous contributions to the user study, and the reviewers for their insightful feedback. 
\end{acks}

\bibliographystyle{ACM-Reference-Format}
\bibliography{100_References}


\end{document}